\documentstyle{elsart}

\input epsf

\begin{document}

\begin{frontmatter}

\title{
Measurement of the radial distribution of Cherenkov light
generated by TeV $\gamma$-ray air showers}

\author[1]{F.A.~Aharonian},
\author[2]{A.G.~Akhperjanian},
\author[3,4]{J.A.~Barrio},
\author[1,9]{K.~Bernl\"ohr},
\author[4]{J.J.G.~Beteta},
\author[6]{H. Bojahr},
\author[4]{J.L.~Contreras},
\author[4]{J.~Cortina},
\author[1]{A.~Daum},
\author[5]{T.~Deckers},
\author[3,4]{J.~Fernandez},
\author[4]{V.~Fonseca},
\author[1]{A.~Fra\ss},
\author[4]{J.C.~Gonzalez},
\author[7]{G.~Heinzelmann},
\author[1]{M.~Hemberger},
\author[1]{G.~Hermann},
\author[1]{M.~He\ss},
\author[1]{A.~Heusler},
\author[1]{W.~Hofmann},
\author[6]{H.~Hohl},
\author[2]{I.~Holl},
\author[7]{D.~Horns},
\author[1,2]{R.~Kankanyan},
\author[3]{M.~Kestel},
\author[5]{O.~Kirstein},
\author[1]{C.~K\"ohler},
\author[1]{A.~Konopelko},
\author[3]{H.~Kornmayer},
\author[3]{D.~Kranich},
\author[1]{H.~Krawczynski},
\author[1]{H.~Lampeitl},
\author[7]{A.~Lindner},
\author[3]{E.~Lorenz},
\author[6]{N.~Magnussen},
\author[6]{H.~Meyer},
\author[3,4,2]{R.~Mirzoyan},
\author[4]{A.~Moralejo},
\author[4]{L.~Padilla},
\author[1]{M.~Panter},
\author[3,6]{D.~Petry},
\author[3]{R.~Plaga},
\author[7]{J.~Prahl},
\author[3]{C.~Prosch},
\author[1]{G.~P\"uhlhofer},
\author[5]{G.~Rauterberg},
\author[6]{W.~Rhode},
\author[7]{A.~R\"ohring},
\author[5]{M.~Samorski},
\author[4]{J.A.~Sanchez},
\author[7]{D.~Schmele},
\author[6]{F.~Schr\"oder},
\author[5]{W.~Stamm},
\author[1]{M.~Ulrich},
\author[1]{H.J.~V\"olk},
\author[6]{B.~Wiebel-Sooth},
\author[1]{C.A.~Wiedner},
\author[5]{M.~Willmer},
\author[1]{H.~Wirth}
 
\collab{HEGRA Collaboration}

\address[1]{Max-Planck-Institut f\"ur Kernphysik, P.O. Box 103980,
        D-69029 Heidelberg, Germany}
\address[2]{Yerevan Physics Institute, Yerevan, Armenia}
\address[3]{Max-Planck-Institut f\"ur Physik, F\"ohringer Ring 6,
        D-80805 M\"unchen, Germany}
\address[4]{Facultad de Ciencias Fisicas, Universidad Complutense,
         E-28040 Madrid, Spain}
\address[5]{Universit\"at Kiel, Inst. f\"ur Kernphysik,
       Olshausenstr.40, D-24118 Kiel, Germany}
\address[6]{BUGH Wuppertal, Fachbereich Physik, Gau\ss str.20,
        D-42119 Wuppertal, Germany}
\address[7]{Universit\"at Hamburg, II. Inst. f\"ur Experimentalphysik,
       Luruper Chaussee 149, D-22761 Hamburg, Germany}
\address[9]{Now at Forschungszentrum Karlsruhe, P.O. Box 3640, 76021 Karlsruhe}
\address[8]{Now at Department of Physics University of Leeds, 
       Leeds LJ2 9JT, UK}

\begin{abstract} 
Using air showers induced by TeV $\gamma$-rays from Mrk 501,
the radial distribution of Cherenkov light is investigated.
The shower geometry is reconstructed from the stereoscopic
shower images obtained with the telescopes of the HEGRA IACT
system. With the core position known to better than 10~m, the
light yield as a function of the distance to the shower axis
is derived by comparing event-by-event the image intensities measured 
in the different telescopes. We observe a change in the shape 
of the light pool with shower energy, and with zenith angle.
Data are well reproduced by Monte-Carlo air shower simulations.
\end{abstract}

\end{frontmatter}

\section{Introduction}

Over the last decade, imaging atmospheric Cherenkov telescopes
(IACTs) have emerged as the prime instrument for the detection
of cosmic $\gamma$-rays in the TeV energy regime \cite{review}. Both galactic
and extragalactic sources of such $\gamma$-rays have been firmly
established, and have been identified with pulsars, supernova
remnants, and active galactic nuclei. Going beyond the existence
proof for different classes of $\gamma$-ray sources, interests
are increasingly turning towards precise measurements of the flux
and of the energy spectra, and the search for a break or cutoff
in the spectra.

Precise measurements of flux and spectrum with the IACT technique
represent a non-trivial challenge. Unlike particle detectors 
used in high-energy-physics experiments or flown on balloons
or satellites, Cherenkov telescopes cannot be calibrated in a
test beam. Their energy calibration and response function has to
be derived indirectly, usually relying heavily on Monte Carlo
simulations. In addition, conventional single IACTs do not allow
to unambiguously reconstruct the full geometry of an air shower, i.e.,
its direction in space and its core location; this lack of 
constraints make consistency checks between data and simulation more
difficult.

The stereoscopic observation of air showers with multiple telescopes,
as pioneered in the HEGRA system of Cherenkov telescopes
\cite{hegra_system}, solves
the latter problem. With two telescopes, the shower geometry is fully
determined. With three or more telescopes, the geometry is overdetermined
and one can measure resolution functions etc. \cite{wh_kruger}.
Angular resolution and energy resolution is improved compared to a 
single telescope. The stereoscopic reconstruction of air showers
also allows a more detailed study of shower properties.

The analysis presented in the following concentrates on one feature
of $\gamma$-ray induced air showers which is central to the reconstruction
of shower energies, namely the distribution of photon intensity in the
Cherenkov light pool, as a function of the distance to the shower core.
In principle, the distribution of Cherenkov light can be calculated 
from first principles, starting from the shower evolution governed by
Quantum Electro Dynamics (QED), 
followed by the well-understood emission of Cherenkov light,
and its propagation through the atmosphere. The relevant atmospheric
parameters are quite well known and parameterized
(see, e.g., \cite{standard_atmo,modtran}). Nevertheless, early simulations showed
significant differences between simulation codes \cite{early_sim}.
These discrepancies can be traced to differences in the assumptions and in the
simplifications which are unavoidable to limit the processor time
required to generate a representative sample of air showers. More
recently, simulation codes seem to have converged 
(see, e.g., \cite{recent_sim}), and
agree reasonably well among each other. Nevertheless, the experimental
verification of this key input to the interpretation of IACT data seems
desirable. In the past, experimental results concerning the distribution
of Cherenkov light in air showers were mainly limited to hadron-induced showers
of much higher energies.

The study of the distribution of Cherenkov light in TeV $\gamma$-ray
showers was carried out using the HEGRA system of IACTs,
based on the extensive sample of $\gamma$-rays detected from the
AGN Mrk 501 \cite{501_paper}. The Mrk 501 $\gamma$-ray sample
combines high statistics with a very favorable ratio of signal to
cosmic-ray background.
The basic idea is quite simple: the shower direction and core location
is reconstructed based on the different views of the shower. One then selects
showers of a given energy and plots the light yield observed in the 
telescopes as a function of the distance to the shower core. 
For this event selection, one should not
use the standard procedures for energy reconstruction
\cite{501_paper,wh_kruger}, since these procedures already assume a 
certain distribution of the light yield. Instead, a much simpler -- and
bias-free -- method
is used to select events of a given energy: one uses a sample of
events which have
their core at a fixed distance $d_i$ (typically around 100~m)
from a given telescope $i$, 
and which generate
a fixed amount of light $a_i$ in this telescope. Located on a circle
around telescope $i$, these showers cover a wide range in core distance
$r_j$ relative to some second telescope $j$, which in case of the
HEGRA array is located between about 70~m and 140~m from telescope $i$.
The measurement of the light yield $a_j$ in this second telescope
provides with $a_j(r_j)$ the shape of the Cherenkov 
light pool. Lacking an absolute energy scale, this method does
provide the radial dependence, but not the absolute normalization of
the light yield. The determine the distribution of light for pure
$\gamma$-rays, the cosmic-ray background under the Mrk 501 signal
is subtracted on a statistical basis.

The following sections briefly describe the HEGRA IACT system, give
more detail on the Mrk 501 data set and the 
analysis technique, and present and summarize the
results.

\section{The HEGRA IACT system}

The HEGRA IACT system is located on the Canary Island of La Palma, 
at the Observatorio del Roque de los Muchachos
of the Instituto Astrofisico de Canarias,
at a height of about 2200~m asl.
The system will ultimately comprise five identical telescopes,
four of which are arranged in the corners of a square with roughly
100~m side length; the fifth telescope is located 
in the center of the square. Currently, four of the telescopes
are operational in their final form. The fifth telescope -- one
of the corner telescopes -- is equipped with an older camera and 
will be upgraded in the near future; it is not included in the
CT system trigger, and is not used in this analysis. 

The system telescopes have
8.5~m$^2$ mirror area, 
5~m focal length, and 271-pixel cameras with a pixel
size of $0.25^\circ$ and a field of view of $4.3^\circ$. 
The cameras are read out by 8-bit Flash-ADCs, which sample the 
pixel signals at a frequency of 120 MHz. More information on 
the HEGRA cameras is given in \cite{hermann_padua}. The two-level
trigger  requires a coincidence of two neighboring pixels
to trigger a telescope, and a coincidence of at least two
telescope triggers to initiate the readout. The pixel trigger
thresholds were initially set to 10~mV, corresponding to
about 8 photoelectrons, and were later in the 1997 run reduced to
8~mV, resulting in a typical trigger rate of 15~Hz, 
and an energy threshold of the system of about 
500~GeV. An in-depth discussion of the trigger system can be
found in \cite{trigger_paper}.

During data taking, a light pulser is used to regularly monitor
the gain and timing of the PMTs. FADC pedestals and offsets of
the trigger discriminators are followed continuously. Deviations
in telescope pointing are measured and corrected using bright
stars, resulting in a pointing accuracy
of better than $0.01^\circ$ \cite{pointing_paper}.

In the data analysis, a deconvolution procedure is applied to 
the FADC data to generate minimum-length signals, and a signal
amplitude and timing is derived for each pixel
\cite{hess_phd}. With the gain
set to about 1 FADC count per photoelectron, the system provides
a direct linear range of about 200 photoelectrons. For larger signals,
the pulse length as measured by the FADC can be used to recover the
amplitude information, extending the dynamic range to well beyond
500 photoelectrons per pixel. Image pixels are then selected as
those pixels having a signal above a high cut of 6 photoelectrons,
or above cut of 3 photoelectrons if adjacent
to a high pixel. By diagonalizing its `tensor of inertia', the major
and minor axes of the images are determined, and the usual {\em width}
and {\em length} parameters \cite{hillas}. Both the image of the source of
a $\gamma$-ray and the point where the shower axis intersects the
telescope plane fall onto the major axes of the images. From the
multiple views of an air shower provided by the different telescopes,
the shower direction is hence determined by superimposing the images
and intersecting their major axes (see \cite{hegra_system,kohnle_paper}
for details); the typical angular resolution is $0.1^\circ$.
Similarly, the core location is derived. The $\gamma$-ray sample is enhanced
by cuts on the {\em mean scaled width} which is calculated by
scaling the measured {\em widths} of all images to the {\em width} expected
for $\gamma$-ray images of a given image {\em size} and distance
to the shower core \cite{501_paper}.

To simulate the properties and detection characteristics of the
HEGRA IACT system, detailed Monte-Carlo simulations are available,
using either the ALTAI \cite{altai} or the CORSIKA \cite{corsika}
air shower generator, followed by
a detailed simulation of the Cherenkov emission and propagation and
of the detector. These simulations include details 
such as the pulse shapes of the input signals to the pixel trigger
discriminators, or the signal recording using the Flash-ADC system
\cite{telsimu1,telsimu2}. In the following, primarily the ALTAI
generator and the detector simulation \cite{telsimu1} was used.
Samples of simulated showers were available for zenith angles of
$0^\circ$, $20^\circ$, $30^\circ$, and $45^\circ$. Distributions at
intermediate angles were obtained by suitably scaling and/or interpolating
the distributions.

\section{The Mrk501 data set}

The extragalactic VHE $\gamma$-ray source Mrk 501 
\cite{whipple_501_initial,hegra_501_initial} showed in 1997 significant
activity, with peak flux levels reaching up to 10 times the
flux of the Crab nebula (see \cite{501_rome} for a summary
of experimental results, first HEGRA results are given in \cite{501_paper}). 
The telescopes of
the HEGRA IACT system were directed towards Mrk 501 for about
140~h, accumulating a total of about 30000 $\gamma$-ray events
at zenith angles between $10^\circ$ and $45^\circ$. Mrk 501
was typically positioned $0.5^\circ$ off the optical axis of the
telescope,
with the sign of the displacement varying every 20~min. In this
mode, cosmic-ray background can be determined by counting events
reconstructed in an equivalent region displaced from the
optical axis by the same amount, but opposite in direction to the
source region; dedicated off-source runs are no longer required,
effectively doubling the net on-source time.
Given the
angular resolution of about $0.1^\circ$, the separation by 
$1^\circ$ of the on-source and off-source regions is fully
sufficient. The relatively large field of view of the cameras
ensures, on the other hand, that images are reliably reconstructed
even with a source displaced from the center of the camera. 

The Mrk 501 $\gamma$-ray data \cite{501_paper}
have provided the basis for a number of
systematic studies of the properties of the HEGRA telescopes, and of the
characteristics of $\gamma$-ray induced air showers (see, e.g., 
\cite{wh_kruger}).

\begin{figure}[htb]
\begin{center}
\mbox{
\epsfxsize7.0cm
\epsffile{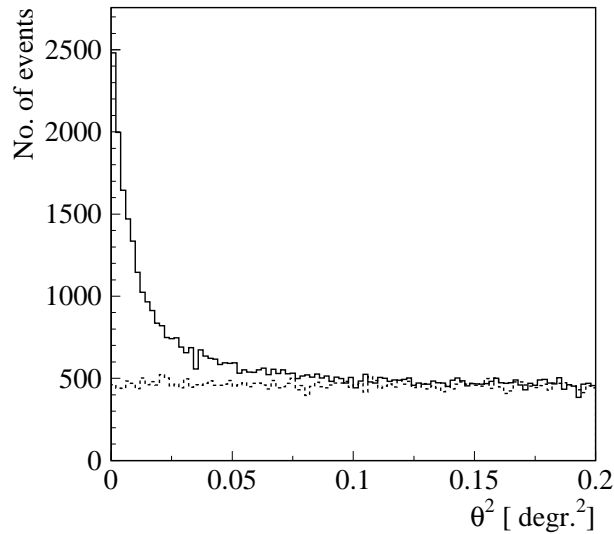}}
\end{center}
\caption
{Distribution in the square of the angle $\theta$ between the 
reconstructed shower axis and the direction
to the source, for events with at least three triggered telescopes.
No cuts on image shapes are applied. The dashed line shows the
distribution for the background region.}
\label{fig_theta2}
\end{figure}

\begin{figure}[htb]
\begin{center}
\mbox{
\epsfxsize7.0cm
\epsffile{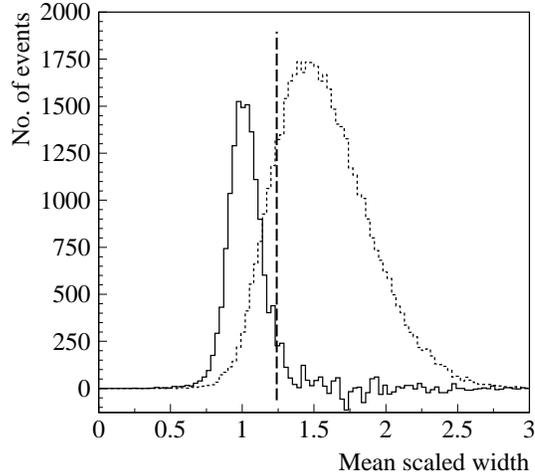}}
\end{center}
\caption
{Distribution in the {\em mean scaled width} for $\gamma$-ray
showers (full line) after statistical subtraction of cosmic rays based on 
the off-source region, and for cosmic rays (dashed).
The dashed line indicates the cut used to select $\gamma$-ray
candidates.}
\label{fig_width}
\end{figure}

\begin{figure}[htb]
\begin{center}
\mbox{
\epsfxsize8.0cm
\epsffile{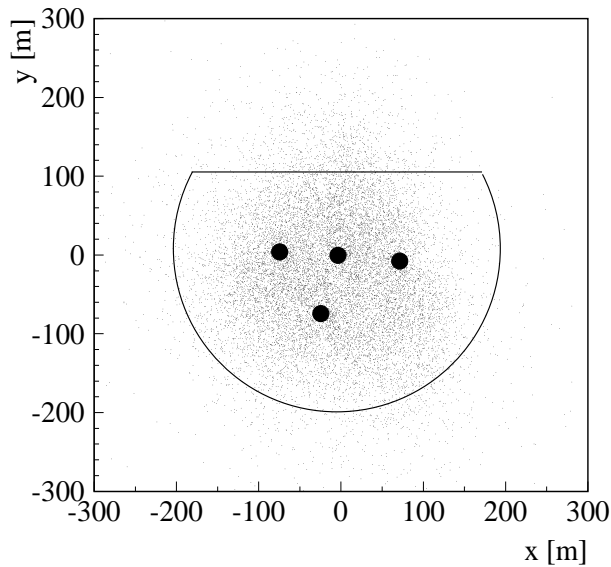}}
\end{center}
\caption
{Distribution of the core locations of events, after the cuts to
enhance the fraction of $\gamma$-rays. Also indicated are the
selection region and the telescope locations.}
\label{fig_core}
\end{figure}

For the following analysis, the data set was cleaned by rejecting
runs with poor or questionable weather conditions, with hardware
problems, or with significant deviations of the trigger rates from
typical values. A subset of events was selected where
at least three telescopes had triggered, and had provided useful
images for the reconstruction of the shower geometry. 
Fig.~\ref{fig_theta2} shows the distribution of reconstructed 
shower axes in the angle $\theta$ relative to the direction towards
Mrk 501. A cut $\theta^2 < 0.05 (^\circ)^2$ was applied to enhance
the $\gamma$-ray content of the sample. 
To further reduce the cosmic-ray background, a loose cut on the
{\em mean scaled width} was used. The distributions in the 
{\em mean scaled width} are shown in Fig.~\ref{fig_width}; events
were selected requiring a value below 1.25; this cut accepts
virtually all $\gamma$-rays.
To ensure that the core location of the events is well reconstructed,
the sample was further restricted to events with a core location
within 200~m from the center of the array (Fig.~\ref{fig_core});
in addition, events with $y_{core} > 100~m$ were rejected,
corresponding to the area near the fifth telescope currently 
not included in the system.
After these cuts, a sample of 11874 on-source events remained, including
a background of 1543 cosmic-ray events, as estimated using the equal-sized
off-source region.

\begin{figure}[htb]
\begin{center}
\mbox{
\epsfxsize7.0cm
\epsffile{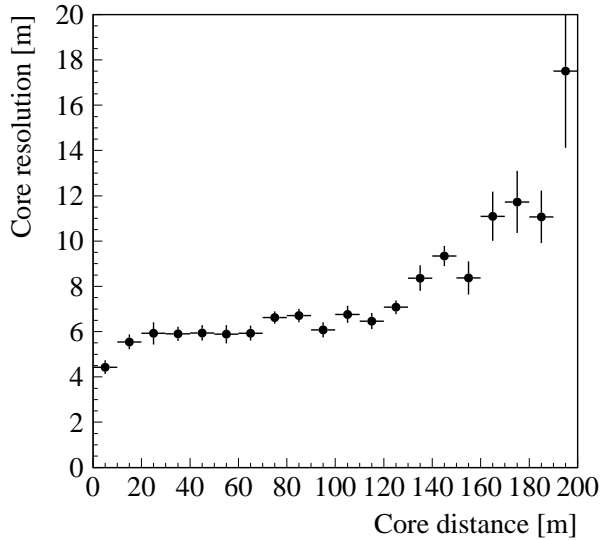}}
\end{center}
\caption
{Resolution in the core position as a function of the distance
between the shower core and the central telescope, as determined
from Monte Carlo simulations of $\gamma$-ray showers with
energies between 1 and 2 TeV. The resolution is defined by
fitting a Gaussian to the distribution of differences between the true and
reconstructed coordinates of the shower impact point, projected
onto the $x$ and $y$ axes of the coordinate system. Due to slight
non-Gaussian tails, the rms widths of the distributions are about
20\% larger.}
\label{fig_coreres}
\end{figure}

For such a sample of events at TeV energies, 
the core location is measured with a
precision of about 6~m to 7~m for events with cores within a 
distance up to 100~m from the central telescope; for larger
distances, the resolution degrades gradually, due to
the smaller angles between the different views,
and the reduced image {\em size} (see Fig.~\ref{fig_coreres}).

\section{The shape of the Cherenkov light pool for $\gamma$-ray
events}

Using the technique described in the introduction, the intensity
distribution in the Cherenkov light pool can now simply be traced
by selecting events with the shower core at given distance $r_i$ from
a `reference' 
telescope $i$ and with a fixed image {\em size} $a_i$, and plotting the
mean amplitude $a_j$ of telescope $j$ as a function of $r_j$.
However, in this simplest form, the procedure is not very practical,
given the small sample of events remaining after such additional
cuts. To be able to use a larger sample of events, one has to
\begin{itemize}
\item select events with $a_i$ in a certain range, $a_{min} < a_i 
< a_{max}$, and plot $a_j/a_i$ vs $r_j$, assuming that the shape of
the light pool does not change rapidly with energy, and that one
can average over a certain energy range
\item repeat the measurement of $a_j(r_j)/a_i$ for different (small) bins 
in $r_i$, and combine these measurements after normalizing the distributions
at some fixed distance
\item Combine the results obtained for different pairs of telescopes $i,j$.
\end{itemize}
Care has to be taken not to introduce a bias due to the trigger
condition. For example, one has to ensure that the selection
criterion of at least three triggered telescopes is fulfilled regardless
of whether telescope $j$ has triggered or not, otherwise the selection
might enforce a minimum image {\em size} in telescope $j$. 

To avoid truncation of images by the border of the camera, only images
with a maximum distance of $1.5^\circ$ between the image centroid and
the camera center were included, leaving a $0.6^\circ$ margin to
the edge of the field of view. Since 
the image of the source if offset by $0.5^\circ$ from the camera 
center, a maximum distance of $2.0^\circ$ is possible between the source
image and the centroid of the shower image.

\begin{figure}[htb]
\begin{center}
\mbox{
\epsfxsize11.0cm
\epsffile{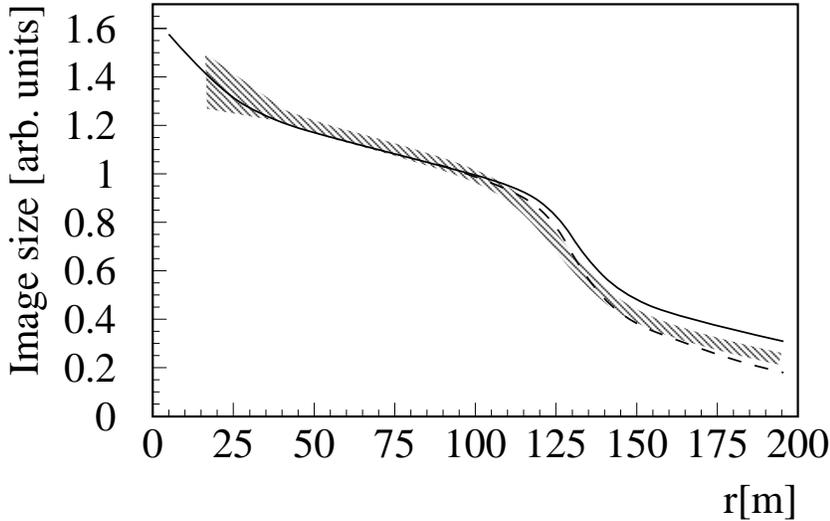}}
\end{center}
\caption
{Radial distribution of Cherenkov light for TeV $\gamma$-ray
showers, for unrestricted aperture of the photon detector (full line),
for a $2^\circ$ aperture (dashed), and
including the full camera simulation and image processing (shaded).
The curves are normalized at $r \approx $100~m.}
\label{fig_pool}
\end{figure}

Even after these selections, the comparison between data and shower models
is not completely straight forward. One should not, e.g., simply compare
data to the predicted photon flux at ground level since
\begin{itemize}
\item as is well known, the radial dependence
of the density of Cherenkov light depends on the solid angle over which
the light is collected, i.e., on the field of view of the camera
\item the experimental resolution in the
reconstruction of the shower core position causes a 
certain smearing, which is visible in particular near the break 
in the light distribution
at the Cherenkov radius
\item the selection of image pixels using the tail cuts results in a
certain loss of photons; this loss is the more significant the lower
the intensity in the image is, and the more diffuse the image is.
\end{itemize}
While the distortion in the measured radial distribution of Cherenkov
light due to the latter two effects is relatively modest (see
Fig.~\ref{fig_pool}), a detailed
comparison with Monte Carlo should take these effects into account by
processing Monte-Carlo generated events using the same procedure as
real data, i.e., by plotting the distance to the reconstructed core
position rather than the true core position, and by applying the same
tail cuts etc.  

\begin{figure}[p]
\begin{center}
\mbox{
\epsfysize18.0cm
\epsffile{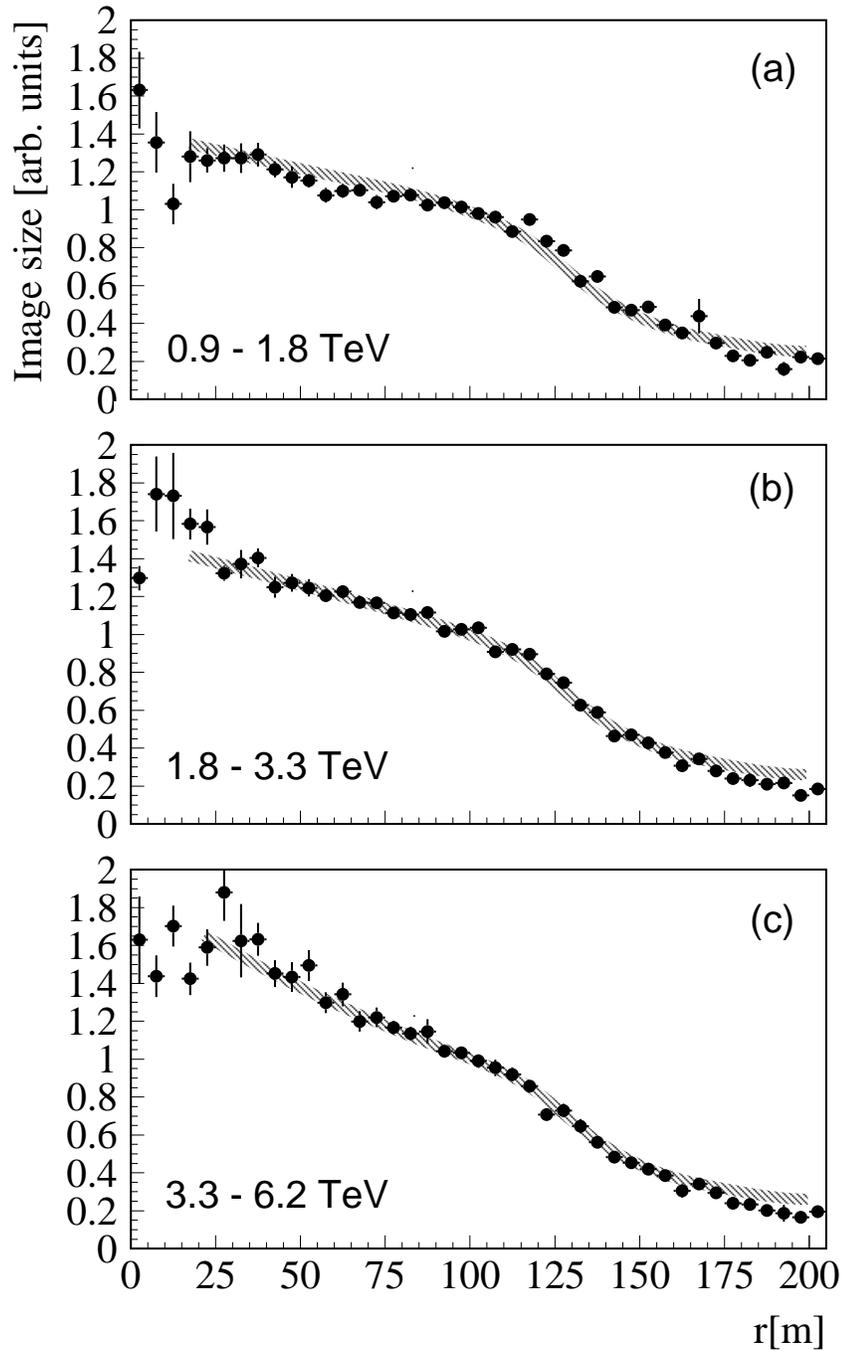}}
\end{center}
\caption
{Light yield as a function of shower energy, for image {\em size} in 
the reference telescope between 100 and 200 photoelectrons (a),
200 and 400 photoelectrons (b), and 400 to 800 photoelectrons (c).
Events were selected 
with a distance range between 50~m and 120~m from the reference telescope,
for zenith angles between $10^\circ$ and $15^\circ$.
The shaded bands indicate the Monte-Carlo results.
The distributions are normalized at $r \approx 100$~m. Only 
statistical errors are shown.}
\label{fig_dat2}
\end{figure}

For a first comparison between data and simulation,
showers from the zenith (zenith angle between
$10^\circ$ and $15^\circ$) were selected. 
The range of distances $r_i$ from the shower core 
to the reference telescope was restricted to the plateau region
between 50~m and 120~m. Smaller
distances were not used because of the large fluctuations of image
{\em size} close to the shower core, and larger distances were excluded
because of the relatively steep variation of light yield with 
distance. The showers were further selected on an amplitude in the `reference'
telescope $i$ between 100 and 200 photoelectrons, corresponding to
a mean energy of about 1.3~TeV. 
Contamination of the Mrk 501 on-source data sample by cosmic
rays  was subtracted using an off-source region displaced from
the optical axis by the same amount as the source, but in
the opposite direction. The measured radial distribution
(Fig.~\ref{fig_dat2}(a))
shows the expected features: a relatively flat plateau out to distances
of 120~m, and a rapid decrease in light yield for larger distances.

The errors given in the Figure are purely statistical. To estimate the
influence of systematic errors, one can look at the consistency of
the data for different ranges in distance $r_i$ to the `reference' 
telescope, one can compare results for different telescope combinations,
and one can study the dependence on the cuts applied. Usually,
the different data sets were consistent to better than $\pm 0.05$ units;
systematic effects certainly do not exceed a level of $\pm 0.1$ units. 
Within these
errors, the measured distribution is  reasonably well reproduced
by the Monte-Carlo
simulations.

\begin{figure}[p]
\begin{center}
\mbox{
\epsfysize20.0cm
\epsffile{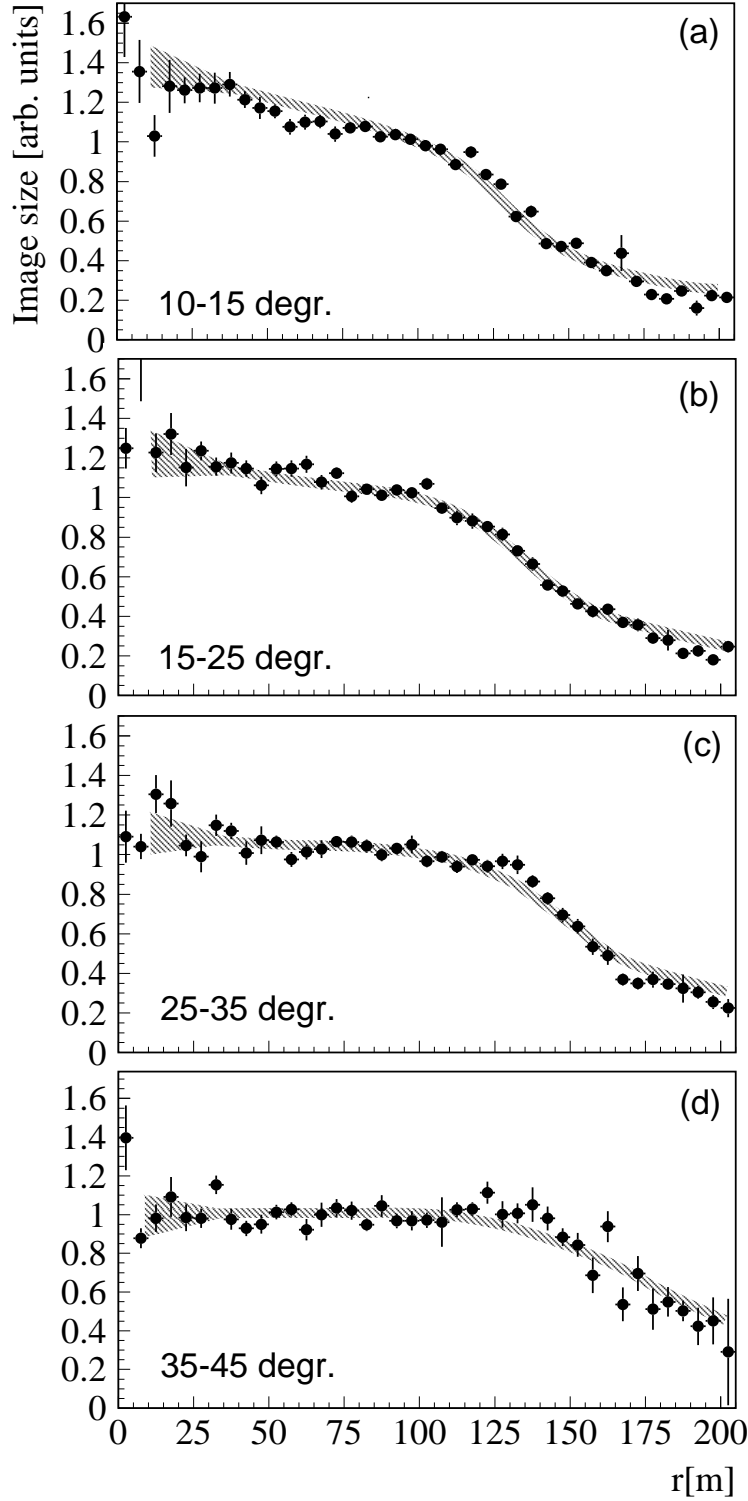}}
\end{center}
\caption
{Light yield as a function of core distance, for zenith angles between
$10^\circ$ and $15^\circ$ (a), $15^\circ$ and $25^\circ$ (b), $25^\circ$ and
$35^\circ$ (c), and $35^\circ$ and $45^\circ$ (d). Events were selected 
with a distance range between 50~m and 120~m from the reference telescope,
and an image {\em size} between 100 and 200 photoelectrons in the reference
telescope. 
The shaded bands indicate the Monte-Carlo results.
The distributions are normalized at $r \approx 100$~m.
Only statistical errors are shown.}
\label{fig_dat3}
\end{figure}

Shower models predict that the distribution
of light intensity varies (slowly) with the shower
energy and with the zenith angle. Fig.~\ref{fig_dat2} compares the
distributions obtained for different {\em size} ranges $a_i$ of
100 to 200, 200 to 400, and 400 to 800 photoelectrons at distances
between 50~m and 120~m, corresponding
to mean shower energies of about 1.3, 2.5, and 4.5 TeV, respectively.
We note that the intensity close to the shower core increases with
increasing energy. This component of the Cherenkov light is generated
by penetrating particles near the shower core. Their number grows
rapidly with increasing shower energy, and correspondingly decreasing
height of the shower maximum. The increase in the mean light intensity 
at small distances from the shower core is primarily caused by
long tails distribution of image {\em sizes} towards large {\em size}; the
median {\em size} is more or less constant.
The observed trends are well reproduced by the
Monte-Carlo simulations.

The dependence on zenith angle is
illustrated in Fig.~\ref{fig_dat3}, where zenith angles between 
$10^\circ$ and $15^\circ$, $15^\circ$ and $25^\circ$, $25^\circ$ and
$35^\circ$, and $35^\circ$ and $45^\circ$ are compared. Events were
again selected for an image {\em size} in the `reference' telescope
between 100 and 200 photoelectrons, in a distance range of 50~m to 
120~m \footnote{Core
distance is always measured in the plane perpendicular to the shower
axis}. The corresponding 
mean shower energies for the four ranges in zenith angle are about 
1.3~TeV, 1.5~TeV, 2~TeV, and 3~TeV.
For increasing zenith angles, the distribution of Cherenkov light
flattens for small radii, and the diameter of the light pool
increases. Both effects are expected, since for larger zenith
angles the distance between the telescope and the shower maximum
grows, reducing the number of penetrating particles, and resulting
in a larger Cherenkov radius. The simulations  properly account for 
this behaviour.

\begin{figure}[tb]
\begin{center}
\mbox{
\epsfxsize7.0cm
\epsffile{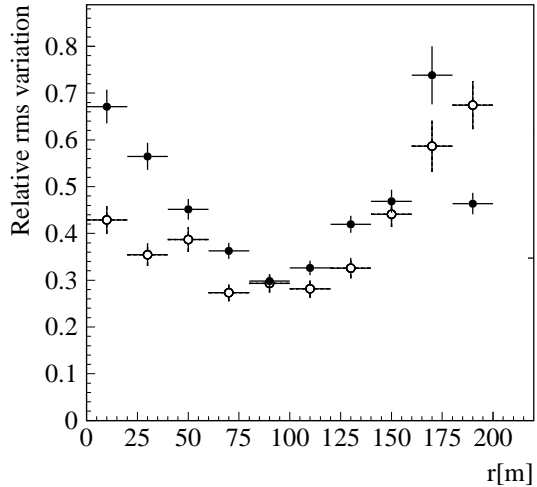}}
\end{center}
\caption
{Relative variation in the {\em size} ratio $a_j/a_i$ as a function
of $r_j$, for $r_i$ in the range 50~m to 120~m, and for image {\em size}
in the `reference' telescope between 100 and 200 photoelectrons.
Full circles refer to zenith angles between $10^\circ$ and $15^\circ$, 
open circles to zenith angles between $25^\circ$ and $35^\circ$.}
\label{fig_rms}
\end{figure}
It is also of some interest to consider the fluctuations of
image {\em size}, $\Delta(a_j/a_i)$.
Fig.~\ref{fig_rms} shows the relative rms fluctuation in the
{\em size} ratio, as a function of $r_j$, for small ($10^\circ$ to
$15^\circ$) and for larger ($25^\circ$ and $35^\circ$) zenith
angles. The fluctuations are minimal near the Cherenkov radius;
they increase for larger distances, primarily due to the smaller
light yield and hence larger relative fluctuations in the number
of photoelectrons. In particular for the small zenith angles,
the fluctuations also increase for small radii, reflecting the
large fluctuations associated with the penetrating tail of the
air showers. For larger zenith angles, this effect is much reduced,
since now all shower particles are absorbed well above the telescopes;
more detailed studies show that already zenith angles of $20^\circ$
make a significant difference. 

\section{Summary}

The stereoscopic observation of $\gamma$-ray induced air showers
with the HEGRA Cherenkov telescopes allowed for the first time
the measurement of the light distribution in the Cherenkov light 
pool at TeV energies, providing a consistency check of one of the
key inputs for the calculation of shower energies based on the 
intensity of the Cherenkov images. The light distribution shows a
characteristic variation with shower energy and with zenith angle.
Data are well reproduced by the Monte-Carlo
simulations.

\section*{Acknowledgements}

The support of the German Ministry for Research 
and Technology BMBF and of the Spanish Research Council
CYCIT is gratefully acknowledged. We thank the Instituto
de Astrofisica de Canarias for the use of the site and
for providing excellent working conditions. We gratefully
acknowledge the technical support staff of Heidelberg,
Kiel, Munich, and Yerevan.

\end{document}